# Superconductivity in the Medium-Entropy/High-Entropy Re-based Alloys with a Non-Centrosymmetric $\alpha$-Mn Lattice

*Kuan Li*[a], *Longfu Li*[a], *Lingyong Zeng*[a,b], *Yucheng Li*[a], *Rui Chen*[a], *Peifeng Yu*[a], *Kangwang Wang*[a], *Zaichen Xiang*[a], *Tian Shang*[c], *Huixia Luo*[adef*]

[a]School of Materials Science and Engineering, Sun Yat-sen University, No. 135, Xingang Xi Road, Guangzhou, 510275, P. R. China

[b]Device Physics of Complex Materials, Zernike Institute for Advanced Materials, University of Groningen, Nijenborgh 4, 9747 AG Groningen, The Netherlands

[c]Key Laboratory of Polar Materials and Devices (MOE), School of Physics and Electronic Science, East China Normal University, Shanghai 200241, China

[d]State Key Laboratory of Optoelectronic Materials and Technologies, Sun Yat-sen University, No. 135, Xingang Xi Road, Guangzhou, 510275, P. R. China

[e]Key Lab of Polymer Composite & Functional Materials, Sun Yat-sen University, No. 135, Xingang Xi Road, Guangzhou, 510275, P. R. China

[f]Guangdong Provincial Key Laboratory of Magnetoelectric Physics and Devices, Sun Yat-sen University, No. 135, Xingang Xi Road, Guangzhou, 510275, P. R. China

*Corresponding author/authors complete details (Telephone; E-mail:) (+86)-2039386124; E-mail address: luohx7@mail.sysu.edu.cn;

## Abstract

Medium/high-entropy alloys (MEAs-HEAs) and rhenium-based compounds with a non-centrosymmetric (NC) structure have received a lot of attention for offering a fertile soil in search for unconventional superconductivity. Here, five previously unreported NC Re-based MEA-HEA superconductors with an $\alpha$-Mn lattice are successfully synthesized, with their superconducting transition temperatures ($T_c$s) ranging from 4 to 5 K. An increase in the superconducting transition temperature ($T_c$) can be achieved by modulating the valence electron count (VEC) through compositional adjustments. Magnetization measurements confirm that all the synthesized Re-based MEA-HEAs are bulk type-II superconductors. Specific heat analysis reveals that the superconducting state of these HEAs can be well described by a single-gap *s*-wave model. Our results show that the Kadowaki-Woods ratio (KWR) of these $\alpha$-Mn MEA/HEA superconductors are close to the typical value of heavy fermion compounds, suggesting the existence of strong electronic correlation. These findings provide promising material platforms to study the role of high disorder in the origin of superconductivity in the NC MEAs-HEAs.

## Introduction

Superconductivity (SC) is a fascinating phenomenon in condensed matter physics that has been attracting researchers around the world due to its unique properties and potential applications [1-3]. In recent years, the investigation of superconducting properties in medium- and high-entropy alloys (MEAs and HEAs) has garnered significant attention, particularly following the groundbreaking discovery of SC in body-centered cubic (BCC) HEAs in 2014 [4-13]. Up to date, many MEA-HEA superconductors have been discovered, which exhibit unconventional electronic structures and intriguing superconducting properties, such as topological bands [2, 14, 15], resistance to high pressure and acid/alkali [7, 16], and strong electron-phonon coupling [17-19].

Usually, MEA-HEA superconductors adopt primary simple structures, such as BCC, face-centered cubic (FCC), hexagonal close-packed (HCP), and CsCl structures [2-8]. While, some superconducting MEA-HEA alloys also crystallize in a complex crystal structure, e.g., the non-centrosymmetric (NC) $\alpha$-Mn structure [13]. Such a structure is derived from a low-temperature isomer of manganese and exhibits a BCC structure [17, 20, 21] with a relatively large unit cell. Compared to the simple BCC and CsCl type structures [22-24] previously reported in superconducting MEAs-HEAs, the NC $\alpha$-Mn structure contains more atoms, with each cell containing 58 atoms. Plenty of previous work indicates that Re-based binary intermetallic and alloy superconductors [ReX (X = transition metal)] with an NC $\alpha$-Mn structure often exhibit unconventional SC [25-30]. Furthermore, superconducting HEAs verify the existence of superconductivity in highly disordered systems lacking regular phonon modes [22]. Compared to BCC and FCC structures, these alloys exhibit extraordinary hardness [31, 32]. $(HfTaWIr)_{1-x}[Re]_x$ and $(HfTaWPt)_{1-x}[Re]_x$ are the first reported $\alpha$-Mn-type HEA superconductors with only 5$d$ elements. Their superconducting transition temperatures ($T_c$s) increase linearly with decreasing lattice parameter $a$ and increasing VEC, which lie between the trend lines of crystalline and amorphous transition metal alloy superconductors [13]. In addition, the upper critical field of the NC $\alpha$-Mn-type $Re_{0.35}Os_{0.35}Mo_{0.10}W_{0.10}Zr_{0.10}$ exceeds the Pauli limiting field, suggesting that unconventional superconducting properties may exist [33]. Therefore, these aforementioned extraordinary features of the MEAs-HEAs, Re-based compounds, and the NC $\alpha$-Mn structure inspire us to design and synthesize new NC $\alpha$-Mn-type MEAs-HEAs and explore their superconductivity as well as mechanism inside these disordered materials.

Here, we present the studies on five newly synthesized Re-based MEA-HEA superconductors with a NC α-Mn Lattice, including $Re_{0.3}Os_{0.3}Ta_{0.1}Nb_{0.1}W_{0.1}Zr_{0.1}$, $Re_{0.35}Os_{0.35}Ta_{0.1}W_{0.1}Zr_{0.1}$, $Re_{0.35}Os_{0.35}Nb_{0.1}W_{0.1}Zr_{0.1}$, $Re_{0.35}Os_{0.35}Mo_{0.1}Ta_{0.1}Zr_{0.1}$, and $Re_{0.4}Os_{0.4}Ta_{0.1}Zr_{0.1}$. These superconductors were synthesized using an arc-melting method. X-ray diffraction, resistivity, magnetic susceptibility, and specific heat measurements are used to characterize the structural and superconducting properties of these new compounds.

## Experiment

Polycrystalline samples of the $Re_{0.3}Os_{0.3}Ta_{0.1}Nb_{0.1}W_{0.1}Zr_{0.1}$, $Re_{0.35}Os_{0.35}Ta_{0.1}W_{0.1}Zr_{0.1}$, $Re_{0.35}Os_{0.35}Nb_{0.1}W_{0.1}Zr_{0.1}$, $Re_{0.35}Os_{0.35}Mo_{0.1}Ta_{0.1}Zr_{0.1}$, and $Re_{0.4}Os_{0.4}Ta_{0.1}Zr_{0.1}$ Re-based MEA-HEA compositions were prepared by an arc-melting method. The raw materials for Re-based superconductors are rhenium from Macklin (99.99%, 325 mesh), osmium from Adamas-beta (99.95%, 200 mesh), molybdenum from Alfa Aesar (99.9%, 250 mesh), niobium from Macklin (99.99%, 325 mesh), tantalum from Alfa Aesar (99.97%, 325 mesh), hafnium from Aladdin (99.9%, 325 mesh), tungsten from Macklin (99.5%, 325 mesh), and Aladdin's zirconium (99.5%, 200 mesh). Five compositions were weighed in their respective molar ratios in a mixture of 300 mg of elemental powders. Then, the samples of the powders were placed in the mortar and thoroughly ground for one hour in in a glove box under an inert gas atmosphere of argon to make a homogeneous mixture. The mixtures were then pressed into pellets using a mold. The pellets were transferred to an electric arc furnace, pumped, and bled argon gas six times to remove the air eventually melted at 0.5 atmospheres in a controlled argon atmosphere. The resulting compounds were then homogenized by melting the ingot several times. All samples had a good metallic luster with negligible loss of mass.

Part of the obtained samples were ground into powders for X-ray diffraction (XRD) measurements. XRD data were collected from 10 ° to 100 ° using a Rigaku MiniFlex (Cu K$\alpha$1 radiation) at a constant scanning speed of 1 °/min. The surface morphology and elemental compositions of the $Re_{0.3}Os_{0.3}Ta_{0.1}Nb_{0.1}W_{0.1}Zr_{0.1}$, $Re_{0.35}Os_{0.35}Ta_{0.1}W_{0.1}Zr_{0.1}$, $Re_{0.35}Os_{0.35}Nb_{0.1}W_{0.1}Zr_{0.1}$, $Re_{0.35}Os_{0.35}Mo_{0.1}Ta_{0.1}Zr_{0.1}$, and $Re_{0.4}Os_{0.4}Ta_{0.1}Zr_{0.1}$ samples were further analyzed by scanning electron microscopy and energy dispersive X-ray spectroscopy (SEM-EDX) with an electron acceleration voltage of 20 kV [2, 15]. Resistivity, magnetization, and heat capacity measurements were conducted employing a physical property measurement system (PPMS) manufactured by Quantum Design. The resistance was determined utilizing the four-probe technique, whereas magnetization and specific heat capacity were assessed on small block samples.

**RESULTS AND ANALYSIS**

The refinement results of the XRD patterns for the $Re_{0.3}Os_{0.3}Ta_{0.1}Nb_{0.1}W_{0.1}Zr_{0.1}$, $Re_{0.35}Os_{0.35}Ta_{0.1}W_{0.1}Zr_{0.1}$, $Re_{0.35}Os_{0.35}Nb_{0.1}W_{0.1}Zr_{0.1}$, $Re_{0.35}Os_{0.35}Mo_{0.1}Ta_{0.1}Zr_{0.1}$, and $Re_{0.4}Os_{0.4}Ta_{0.1}Zr_{0.1}$ samples are presented in **Figure 1a** to **Figure 1e**. The refinements of the XRD patterns indicate that five obtained samples all have single phase and crystallize in a NC α-Mn structure with the space group $I\bar{4}3m$. The refinement parameters $R_{wp}$, $R_p$, $R_e$, and $\chi^2$ for the Rietveld fit of $Re_{0.3}Os_{0.3}Ta_{0.1}Nb_{0.1}W_{0.1}Zr_{0.1}$ are 7.42%, 5.22%, 2.84%, and 6.0833, respectively, suggesting a satisfactory fit of our data. The XRD fitting parameters and unit cell parameters of all MEA-HEA superconductors are summarized in **Table S1**. The **Figure 1f** illustrates the representative sample $Re_{0.35}Os_{0.35}Nb_{0.1}W_{0.1}Zr_{0.1}$ crystal structure in the cubic crystal HEA system. All the atoms are randomly distributed in the lattice. Besides that, according to the definition of mixing entropy $\Delta s_{mix} = -R\sum_{i=1}^{n} c_i \ln c_i$, where $n$ is the number of components, $c_i$ is the atomic fraction, and $R$ is the gas constant, the $\Delta S_{mix}$s for all five superconductors are also presented in **Table S1**. **Table S2** lists the atomic parameters of a representative sample of $Re_{0.4}Os_{0.4}Ta_{0.1}Zr_{0.1}$ superconductor.

In addition, the homogeneity and chemical composition of these Re-based HEAs were characterized by SEM-EDX [2,15]. To ensure the accuracy of the measurement, we tested the EDX of each sample multiple times. The **Table S3** shows the specific data of the average value for each element. From the SEM-EDX elemental mappings, as shown in **Figure S2**, we can see that all elements are evenly distributed and no clusters are seen, indicating that these obtained five Re-based HEAs are in a single solid solution phase. In addition, the calculated components are $Re_{0.31}Os_{0.30}Ta_{0.10}Nb_{0.07}W_{0.10}Zr_{0.12}$, $Re_{0.35}Os_{0.36}Ta_{0.1}W_{0.09}Zr_{0.1}$, $Re_{0.36}Os_{0.36}Nb_{0.06}W_{0.10}Zr_{0.12}$, $Re_{0.34}Os_{0.35}Mo_{0.08}Ta_{0.10}Zr_{0.13}$ and $Re_{0.39}Os_{0.39}Ta_{0.11}Zr_{0.11}$, in which the actual elemental ratios are very close to the design values, indicating that the prepared samples are as expected.

The temperature-dependent resistivity $\rho$(T) of these Re-based HEAs measured between 300 K and 1.8 K are shown in **Figure 2a**. The room temperature resistivities of $Re_{0.3}Os_{0.3}Ta_{0.1}Nb_{0.1}W_{0.1}Zr_{0.1}$, $Re_{0.35}Os_{0.35}Ta_{0.1}W_{0.1}Zr_{0.1}$, $Re_{0.35}Os_{0.35}Nb_{0.1}W_{0.1}Zr_{0.1}$, $Re_{0.35}Os_{0.35}Mo_{0.1}Ta_{0.1}Zr_{0.1}$, and $Re_{0.4}Os_{0.4}Ta_{0.1}Zr_{0.1}$ are 182, 207, 216, 172, and 317 μΩ cm, respectively. However, at much lower temperatures, the resistivities drop dramatically and tend to zero, clearly indicating the presence of a superconducting transition. The residual resistivity ratio (*RRR*) is a key parameter in our research as it indicates the quality of the material under study [34-36]. In general, higher *RRR* values indicate lower levels of impurities and defects, which are essential to ensure structural homogeneity and excellent electrical conductivity in metallic systems [35]. Here, our calculated *RRR* value is the ratio of the resistivity of the samples at 300 K and 6 K, namely $RRR = \rho_{300K}/\rho_{6K}$, which is used to evaluate the purity and conductivity of these five $\alpha$-Mn-type HEA superconductors. The *RRR* values for the $Re_{0.3}Os_{0.3}Ta_{0.1}Nb_{0.1}W_{0.1}Zr_{0.1}$, $Re_{0.35}Os_{0.35}Ta_{0.1}W_{0.1}Zr_{0.1}$, $Re_{0.35}Os_{0.35}Nb_{0.1}W_{0.1}Zr_{0.1}$,

$Re_{0.35}Os_{0.35}Mo_{0.1}Ta_{0.1}Zr_{0.1}$, and $Re_{0.4}Os_{0.4}Ta_{0.1}Zr_{0.1}$ HEA superconductors are 1.07, 1.01, 1.00, 1.04, and 1.05, respectively. Their *RRR* values are comparable to those of the previously reported MEA-HEA compounds [5, 6, 8].

As shown in **Figure 2b,** all five MEA-HEA Re-based superconductors exhibit a sharp transition to the zero-resistivity state. We determined that $T_c^{\rho}$ is a 50% decrease in resistivity, concerning the superconducting transition state value. The obtained $T_c^{\rho}$ of the $Re_{0.3}Os_{0.3}Ta_{0.1}Nb_{0.1}W_{0.1}Zr_{0.1}$, $Re_{0.35}Os_{0.35}Ta_{0.1}W_{0.1}Zr_{0.1}$, $Re_{0.35}Os_{0.35}Nb_{0.1}W_{0.1}Zr_{0.1}$, $Re_{0.35}Os_{0.35}Mo_{0.1}Ta_{0.1}Zr_{0.1}$ and $Re_{0.4}Os_{0.4}Ta_{0.1}Zr_{0.1}$ HEA superconductors are around 4.0(3) K, 4.4(5) K, 4.5(2) K, 4.6(3) K, and 5.0(2) K, respectively. It is worth that the range of drastic changes in resistivity ($\Delta\rho$) occurs in a very narrow range. This indicates that the material is highly pure and well-homogenized. The $T_c^{\rho}$ = 4.9(2) K of $Re_{0.35}Os_{0.35}Mo_{0.1}W_{0.1}Z_{r0.1}$ is marginally higher than the $T_c^{\rho}$ of $Re_{0.35}Os_{0.35}Mo_{0.1}Ta_{0.1}Zr_{0.1}$ ($T_c^{\rho}$ = 4.6(3) K), while lower than that of $Re_{0.4}Os_{0.4}Ta_{0.1}Zr_{0.1}$ ($T_c^{\rho}$ = 5.0(2) K). These superconductors are all type-II superconductors and all crystallize in the $I\bar{4}3m$ space group. Additionally, the VEC of $Re_{0.35}Os_{0.35}Mo_{0.10}W_{0.10}Zr_{0.10}$ is 6.85, which is slightly higher than those of $Re_{0.35}Os_{0.35}Ta_{0.1}W_{0.1}Zr_{0.1}$, $Re_{0.35}Os_{0.35}Nb_{0.1}W_{0.1}Zr_{0.1}$, and $Re_{0.35}Os_{0.35}Mo_{0.1}Ta_{0.1}Zr_{0.1}$. This also confirms the conclusion that the increase in $T_c^{\rho}$ of the $\alpha$-Mn - structured MEA/HEA superconductors can be realized by add the VEC.

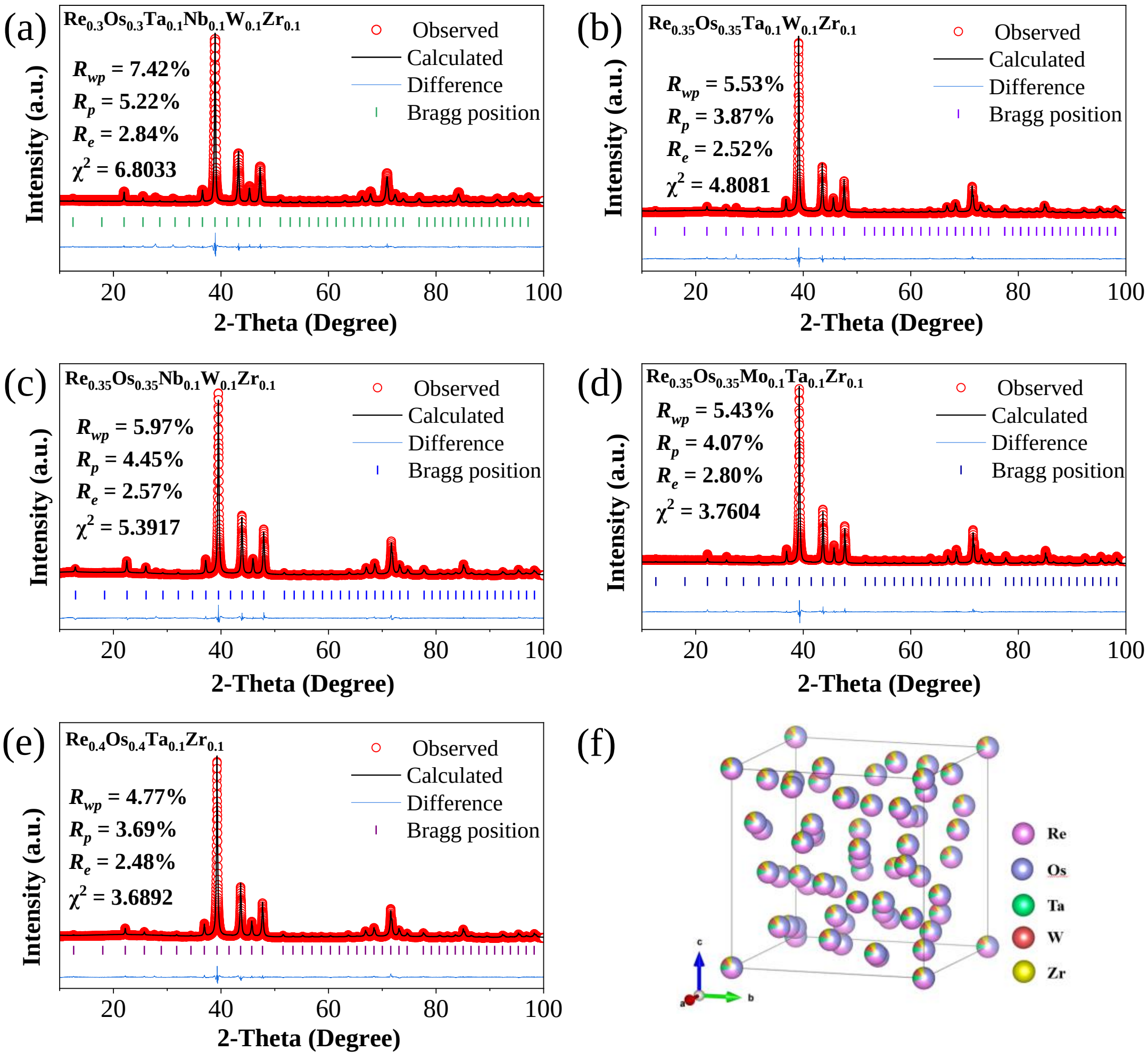

**Figure 1.** (a)-(e) Rietveld refinement profiles of the XRD for $Re_{0.3}Os_{0.3}Ta_{0.1}Nb_{0.1}W_{0.1}Zr_{0.1}$, $Re_{0.35}Os_{0.35}Ta_{0.1}W_{0.1}Zr_{0.1}$, $Re_{0.35}Os_{0.35}Nb_{0.1}W_{0.1}Zr_{0.1}$, $Re_{0.35}Os_{0.35}Mo_{0.1}Ta_{0.1}Zr_{0.1}$, and $Re_{0.4}Os_{0.4}Ta_{0.1}Zr_{0.1}$. (f) Crystal structure of the $Re_{0.35}Os_{0.35}Nb_{.1}W_{0.1}Zr_{0.1}$ HEA superconductor with $I\bar{4}3m$ space group.

**Table 1.** Normal and superconducting-state properties of the five $Re_{0.3}Os_{0.3}Ta_{0.1}Nb_{0.1}W_{0.1}Zr_{0.1}$, $Re_{0.35}Os_{0.35}Ta_{0.1}W_{0.1}Zr_{0.1}$, $Re_{0.35}Os_{0.35}Nb_{0.1}W_{0.1}Zr_{0.1}$, $Re_{0.35}Os_{0.35}Mo_{0.1}Ta_{0.1}Zr_{0.1}$, and $Re_{0.4}Os_{0.4}Ta_{0.1}Zr_{0.1}$ MEA-HEA superconductors.

| **Compound** | $Re_{0.3}Os_{0.3}Ta_{0.1}$ $Nb_{0.1}W_{0.1}Zr_{0.1}$ | $Re_{0.35}Os_{0.35}Ta_{0.1}$ $W_{0.1}Zr_{0.1}$ | $Re_{0.35}Os_{0.35}Nb_{0.1}$ $W_{0.1}Zr_{0.1}$ | $Re_{0.35}Os_{0.35}Mo$ $_{0.1}Ta_{0.1}Zr_{0.1}$ | $Re_{0.4}Os_{0.4}$ $Ta_{0.1}Zr_{0.1}$ |
|---|---|---|---|---|---|
| $T_c^{\rho}$ **(K)** | 4.0(3) | 4.4(5) | 4.5(2) | 4.6(3) | 5.0(2) |
| $T_c^{M}$ **(K)** | 3.8(2) | 4.2(1) | 4.4(3) | 4.5(2) | 4.6(3) |
| $T_c^{HC}$ **(K)** | 3.7(2) | 4.2(3) | 4.3(5) | 4.3(3) | 4.4(3) |
| $\mu_0 H_{c1}(0)$(mT) | 4.22(2) | 5.32(3) | 3.44 (2) | 5.48(3) | 2.76(4) |
| $\mu_0 H_{c2}(0)^{GL}$(T) | 4.91(2) | 5.53(2) | 5.49(2) | 7.06(1) | 7.14(2) |
| $\mu_0 H_{c2}(0)^{WHH}$(T) | 4.31(3) | 5.05(3) | 5.06(4) | 6.08(3) | 6.39(3) |
| $\mu_0 H^{\mathrm{Pauli}}$(T) | 7.44 | 8.18 | 8.37 | 8.56 | 9.3 |
| $\gamma$ (mJ mol$^{-1}$ K$^{-2}$) | 3.314(2) | 3.250(2) | 3.457(2) | 3.304(2) | 3.446(3) |
| $\beta$ (mJ / mol / K$^4$) | 0.048(5) | 0.049(3) | 0.039(3) | 0.039(3) | 0.041(2) |
| $\Theta_D$ ($K$) | 343 | 341 | 368 | 368 | 362 |
| $\Delta C_{el}/\gamma T_c$ | 1.403 | 1.518 | 1.417 | 1.515 | 1.396 |
| $\lambda_{ep}$ | 0.58 | 0.59 | 0.59 | 0.59 | 0.61 |
| $2\Delta_0/k_B T_c$ | 3.49 | 3.63 | 3.51 | 3.63 | 3.48 |
| $\xi_{GL}(0)$ (nm) | 8.2 | 7.7 | 7.7 | 6.8 | 6.8 |

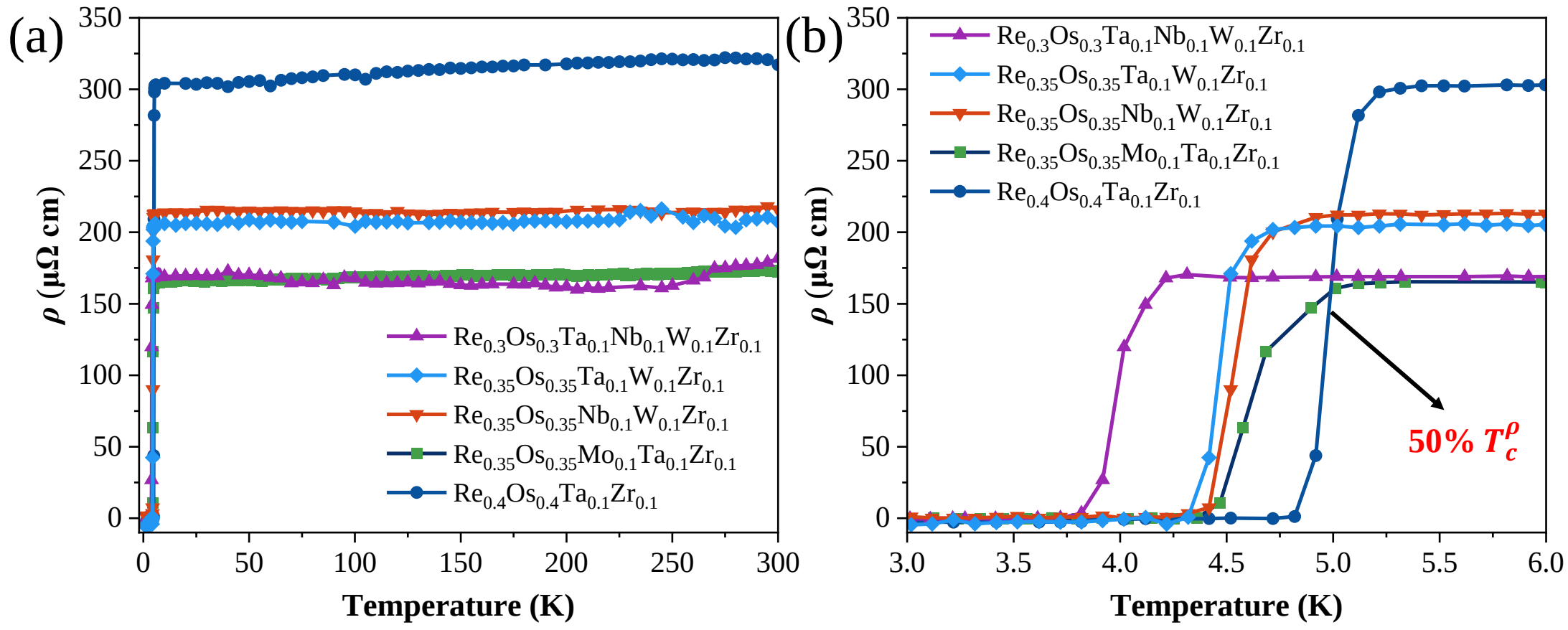

**Figure 2.** (a) The $\rho$(T) measurements of the $Re_{0.3}Os_{0.3}Ta_{0.1}Nb_{0.1}W_{0.1}Zr_{0.1}$, $Re_{0.35}Os_{0.35}Ta_{0.1}W_{0.1}Zr_{0.1}$, $Re_{0.35}Os_{0.35}Nb_{0.1}W_{0.1}Zr_{0.1}$, $Re_{0.35}Os_{0.35}Mo_{0.1}Ta_{0.1}Zr_{0.1}$, and $Re_{0.4}Os_{0.4}Ta_{0.1}Zr_{0.1}$ HEA superconductors in the temperature range from 1.8 K to 300 K. (b) The low-temperature behavior of $\alpha$-Mn-type HEA superconductors in the vicinity of $T_c$.

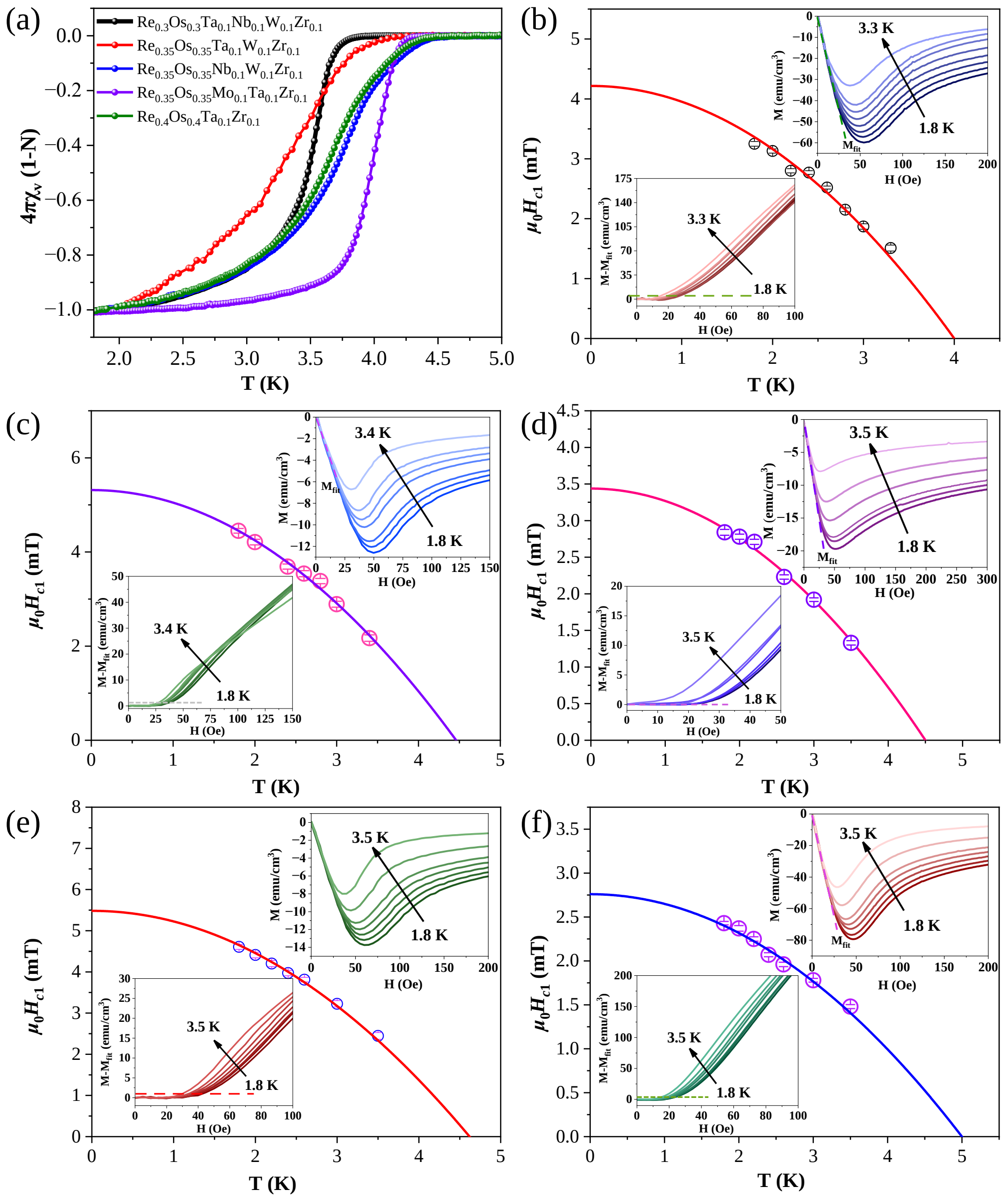

**Figure 3.** (a) The magnetization behavior of the $Re_{0.3}Os_{0.3}Ta_{0.1}Nb_{0.1}W_{0.1}Zr_{0.1}$, $Re_{0.35}Os_{0.35}Ta_{0.1}W_{0.1}Zr_{0.1}$, $Re_{0.35}Os_{0.35}Nb_{0.1}W_{0.1}Zr_{0.1}$, $Re_{0.35}Os_{0.35}Mo_{0.1}Ta_{0.1}Zr_{0.1}$, and $Re_{0.4}Os_{0.4}Ta_{0.1}Zr_{0.1}$ superconductors were measured in a 2 mT magnetic field. (b) - (f) The $\mu_0H_{c1}$ vs temperature for five $\alpha$-Mn-type HEA superconductors. The upper parts of the internal diagrams are magnetization vs applied magnetic field recorded at different temperatures. The lower parts of the internal diagrams are the $M$-$M_{fit}$ curves as a function of the magnetic fields.

To characterize the superconducting volume fraction and lower critical magnetic field of these five Re-based HEA superconductors, we carried the magnetization measurements. The zero-field-cooling (ZFC) magnetic susceptibilities for the Re-based

HEA superconductors in the temperature ranges of 1.8 K to 5 K are shown in **Figure 3a**. The demagnetization $N$ depends on the shape of the sample and the orientation of the sample with respect to the magnetic field [37]. It is determined using the formula - n = $1/(4\pi(1-N))$. **Figure 3a** shows the magnetic susceptibility data corrected with the value of $N$. The resulting diamagnetic signals are very close to the ideal value of -1. The $T_c^M$ s are determined by identifying the intersection points of the linearly extrapolated diamagnetic responses near the superconducting phase transition and the normal-state magnetizations. Strong diamagnetic signals are observed below the temperature points of the $T_c^M$s of 3.8(2) K, 4.2(1) K, 4.4(3) K, 4.5(2) K, and 4.6(3) K, respectively, which further confirm the presence of bulk superconductivity in these Re-based HEA superconductors. To determine lower critical magnetic field ($\mu_0 H_{c1}(0)$), we measured the field-dependent magnetization strengths $M(H)$ of these Re-based HEAs at different temperatures below $T_c$, as shown in the upper insets in **Figure 3b** to **3f**. The magnetization data in the low magnetic field region varies linearly and starts to deviate at the value of the magnetic field. The linear fitting equation for the low magnetic field region is $M_{fit} = e + fH$, where $e$ is the intercept and $f$ is the slope of the line [38]. The lower insets in **Figure 3b** to **3f** show the $M$-$M_{\text{fit}}$ curves for $Re_{0.3}Os_{0.3}Ta_{0.1}Nb_{0.1}W_{0.1}Zr_{0.1}$, $Re_{0.35}Os_{0.35}Ta_{0.1}W_{0.1}Zr_{0.1}$, $Re_{0.35}Os_{0.35}Nb_{0.1}W_{0.1}Zr_{0.1}$, $Re_{0.35}Os_{0.35}Mo_{0.1}Ta_{0.1}Zr_{0.1}$, and $Re_{0.4}Os_{0.4}Ta_{0.1}Zr_{0.1}$ at different temperatures, respectively. The $\mu_0 H_{c1}(\mathrm{T})$ is extracted when the difference between $M$ and $M_{\text{fit}}$ exceeds 1% $M_{\text{max}}$. The obtained $\mu_0 H_{c1}(\mathrm{T})$ values with temperature are shown in **Figure 3b** to **3f**. The dashed lines correspond to the Ginzburg - Landau equation: $\mu_0 H_{c1}(\mathrm{T}) = \mu_0 H_{c1}(0)(1-(T/T_c)^2)$, which yield values of the $\mu_0 H_{c1}(0)$ of $Re_{0.3}Os_{0.3}Ta_{0.1}Nb_{0.1}W_{0.1}Zr_{0.1}$, $Re_{0.35}Os_{0.35}Ta_{0.1}W_{0.1}Zr_{0.1}$, $Re_{0.35}Os_{0.35}Nb_{0.1}W_{0.1}Zr_{0.1}$, $Re_{0.35}Os_{0.35}Mo_{0.1}Ta_{0.1}Zr_{0.1}$, and $Re_{0.4}Os_{0.4}Ta_{0.1}Zr_{0.1}$ are 4.22(2) mT, 5.32(3) mT, 3.44(2) mT, 5.48(3) mT, and 2.76(4) mT, respectively. Low values of the $\mu_0 H_{c1}(0)$ indicate that the material is relatively susceptible to external magnetic field disturbances.

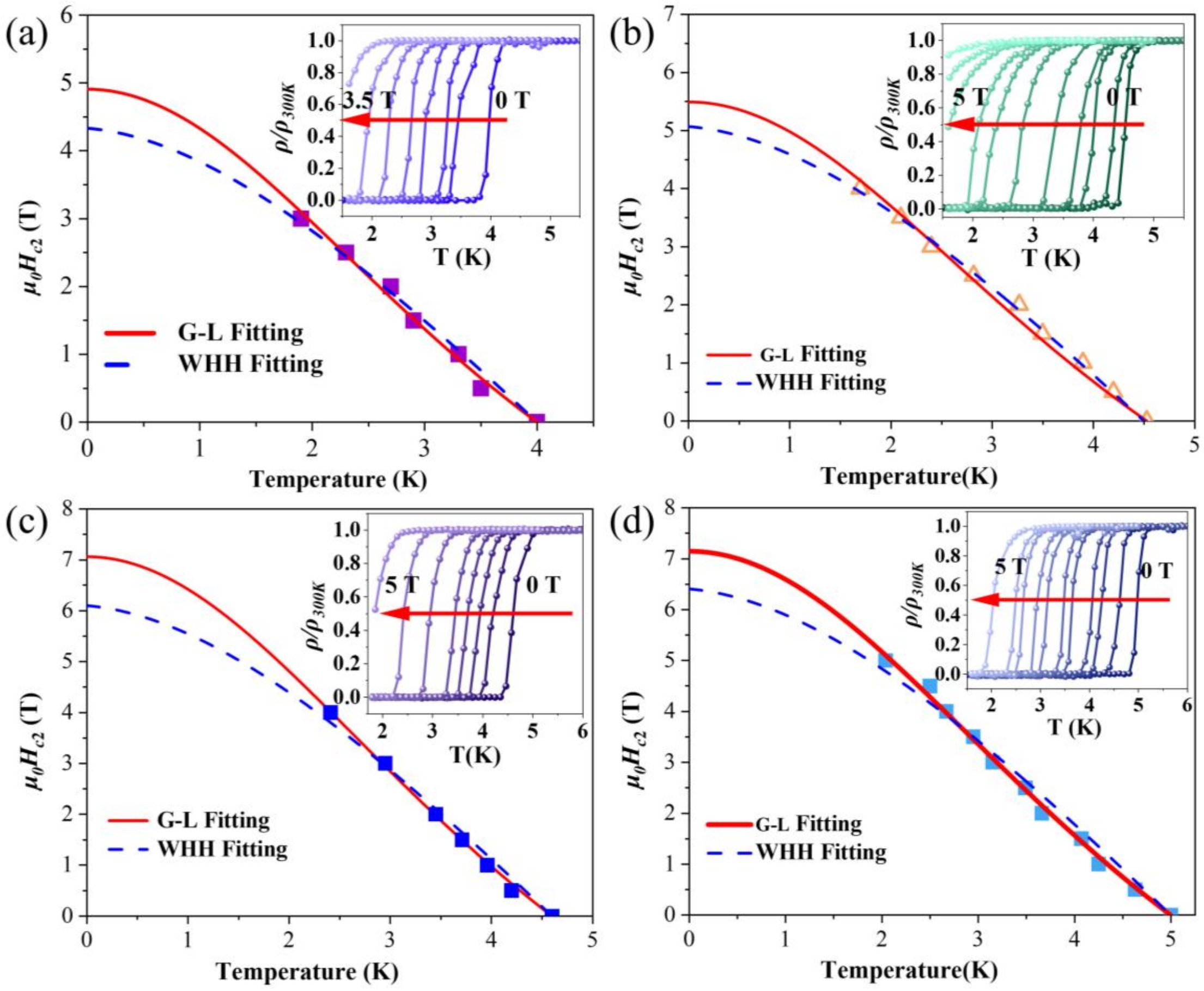

**Figure 4.** (a) - (d) The temperature dependence measurement of the $\mu_0H_{c2}$ for $Re_{0.3}Os_{0.3}Ta_{0.1}Nb_{0.1}W_{0.1}Zr_{0.1}$, $Re_{0.35}Os_{0.35}Nb_{0.1}W_{0.1}Zr_{0.1}$, $Re_{0.35}Os_{0.35}Mo_{0.1}Ta_{0.1}Zr_{0.1}$, and $Re_{0.4}Os_{0.4}Ta_{0.1}Zr_{0.1}$. The $\mu_0H_{c2}$-T phase diagram of HEA superconductors fitted with the GL and WHH equation. The insets show resistive transition under different magnetic fields.

Understanding and investigating the upper critical field of superconductors is crucial for revealing superconducting properties, evaluating material performance, and exploring potential applications of superconductors in high magnetic field environments. Therefore, the electrical resistivity of these Re-based superconductors were systematically measured under various fixed magnetic fields. The insets of **Figure 4a - 4d** show the resistivity of $Re_{0.3}Os_{0.3}Ta_{0.1}Nb_{0.1}W_{0.1}Zr_{0.1}$, $Re_{0.35}Os_{0.35}Nb_{0.1}W_{0.1}Zr_{0.1}$, $Re_{0.35}Os_{0.35}Mo_{0.1}Ta_{0.1}Zr_{0.1}$, and $Re_{0.4}Os_{0.4}Ta_{0.1}Zr_{0.1}$ in the magnetic field, respectively. Meanwhile, **Figure S2** depicts the resistivity of $Re_{0.35}Os_{0.35}Ta_{0.1}W_{0.1}Zr_{0.1}$ in the 0 - 4.5 T magnetic field range. As the magnetic field increases, $T_c^{\rho}$ moves towards lower temperatures. The increase in applied magnetic field suppresses $T_c^{\rho}$. The Ginzburg-Landan (GL) formula $\mu_0 H_{c2}(T) = \mu_0 H_{c2}(0) \times \frac{(1-(T/T_c)^2)}{(1+(T/T_c)^2)}$ was used to analyze the upper critical fields. Across the entire temperature range, the GL model fits the experimental data well. The $\mu_0 H_{c2}(0)^{GL}$ estimates for $Re_{0.3}Os_{0.3}Ta_{0.1}Nb_{0.1}W_{0.1}Zr_{0.1}$, $Re_{0.35}Os_{0.35}Ta_{0.1}W_{0.1}Zr_{0.1}$, $Re_{0.35}Os_{0.35}Nb_{0.1}W_{0.1}Zr_{0.1}$, $Re_{0.35}Os_{0.35}Mo_{0.1}Ta_{0.1}Zr_{0.1}$, and $Re_{0.4}Os_{0.4}Ta_{0.1}Zr_{0.1}$ are 4.91(2) T, 5.53(2) T, 5.49(2) T, 7.06(1) T, and 7.14(2)T, respectively. After successfully fitting the experimental data using the GL formula, we further provide a comprehensive analysis of the $\mu_0 H_{c2}(0)$ using the Werthamer-Helfand-Hohenberg (WHH) formula: $\mu_0 H_{c2}(0) = -0.693 T_c (\frac{d\mu_0 H_{c2}}{dT})|_{T = T_c}$. The $T_c$ values at different magnetic fields can be fitted to a straight line. Both the GL and WHH models fit the experimental data well over the entire temperature range as shown in **Figure 4a** and **4b**. The estimated values of $\mu_0 H_{c2}(0)^{WHH}$ for $Re_{0.3}Os_{0.3}Ta_{0.1}Nb_{0.1}W_{0.1}Zr_{0.1}$, $Re_{0.35}Os_{0.35}Ta_{0.1}W_{0.1}Zr_{0.1}$, $Re_{0.35}Os_{0.35}Nb_{0.1}W_{0.1}Zr_{0.1}$, $Re_{0.35}Os_{0.35}Mo_{0.1}Ta_{0.1}Zr_{0.1}$, and $Re_{0.4}Os_{0.4}Ta_{0.1}Zr_{0.1}$ are 4.31(3) T, 5.05(3) T, 5.06(4) T, 6.08(3) T, and 6.39(3) T, respectively. Based on the BCS theory, the Pauling limit field of a superconductor can be described by $\mu_0 H^{\mathrm{Pauli}}$ = 1.86*$T_c$. In this case, the $\mu_0 H^{\mathrm{Pauli}}$ is 7.44 T, 8.18 T, 8.37 T, 8.56 T, and 9.3 T for $Re_{0.3}Os_{0.3}Ta_{0.1}Nb_{0.1}W_{0.1}Zr_{0.1}$, $Re_{0.35}Os_{0.35}Ta_{0.1}W_{0.1}Zr_{0.1}$, $Re_{0.35}Os_{0.35}Nb_{0.1}W_{0.1}Zr_{0.1}$, $Re_{0.35}Os_{0.35}Mo_{0.1}Ta_{0.1}Zr_{0.1}$, and $Re_{0.4}Os_{0.4}Ta_{0.1}Zr_{0.1}$, respectively. The values of $\mu_0 H_{c2}(0)$ obtained by both GL and WHH models are smaller than those of $\mu_0 H^{\mathrm{Pauli}}$, indicating that orbit pairing breaking is dominant in these superconductors.

After observing the Meissner effect and zero resistivity, we measured the heat capacity of the HEA superconductor with an applied magnetic field, which further confirms the bulk SC, as shown in **Figure 5a-5e**. Further insight into the superconducting nature of five NC $\alpha$-Mn-structure HEA superconductors can be gained by detailed analysis of the specific heat data. The electronic specific heat ($C_{el}$) as a function of temperature is shown in **Figure 5f - 5j**. We analyze the data using the so-called $\alpha$ model [17]. The $\alpha$ model still assumes that the superconducting gap is completely isotropic, but allows for variations in the coupling constant $\alpha = \Delta_0/k_BT_c$, where $\Delta_0$ is the size of the gap at 0 K. The $T_c^{HC}$ in 0 T heat capacity corresponds to the superconducting state appearance of 3.7(2) K for $Re_{0.3}Os_{0.3}Ta_{0.1}Nb_{0.1}W_{0.1}Zr_{0.1}$, 4.2(3) K for $Re_{0.35}Os_{0.35}Ta_{0.1}W_{0.1}Zr_{0.1}$, 4.3(5) K for $Re_{0.35}Os_{0.35}Nb_{0.1}W_{0.1}Zr_{0.1}$, 4.3(3) K for $Re_{0.35}Os_{0.35}Mo_{0.1}Ta_{0.1}Zr_{0.1}$, and 4.4(3) K for $Re_{0.4}Os_{0.4}Ta_{0.1}Zr_{0.1}$, which is in approximately the same agreement as the measurements of resistivity and magnetization strength. These five $\alpha$-Mn-structure HEAs superconducting heat capacity jumps are suppressed under high magnetic fields. In the normal state above $T_c$, the specific heat in the normal state can be fitted by $C_p/T = \gamma + \beta T^2 + \eta T^4$, where $\gamma$ denotes the normal-state electronic specific heat coefficient, $\beta T^2 + \eta T^4$ term represents the phonon contribution. Such fitting has been applied to transition-metal monocarbides (TMMCs) superconductors, Kagome lattice superconductors, and some unconventional superconductors as well [39-41]. The obtained parameters from above analysis are $\gamma$ = 3.314(2) mJ/mol/K$^2$ and $\beta$ = 0.048(5) mJ/mol/K$^4$ for $Re_{0.3}Os_{0.3}Ta_{0.1}Nb_{0.1}W_{0.1}Zr_{0.1}$. The normalized specific heat jumps $\Delta C_{el}/\gamma T_c$ = 1.403 for $Re_{0.3}Os_{0.3}Ta_{0.1}Nb_{0.1}W_{0.1}Zr_{0.1}$.

The Debye temperature ($\Theta_D$) is then estimated using the Debye model and the equation $\Theta_D=(12\pi^4 nR/5\beta)^{1/3}$, where $R$ denotes the gas constant and $n$ denotes the number of atoms in the formulae cell. The estimated $\Theta_D$ values for $Re_{0.3}Os_{0.3}Ta_{0.1}Nb_{0.1}W_{0.1}Zr_{0.1}$ is 343 K. Given $\Theta_D$ and $T_c$, the semiempirical McMillan formula is used: $\lambda_{ep}=\frac{1.04+\mu^* \ln\left(\frac{\Theta_D}{1.45T_c}\right)}{\left(1-1.62\mu^*\right)\ln\left(\frac{\Theta_D}{1.45T_c}\right)-1.04}$ to calculated the electron-phonon coupling constant that is $\lambda_{ep}$ = 0.58 for $Re_{0.3}Os_{0.3}Ta_{0.1}Nb_{0.1}W_{0.1}Zr_{0.1}$. Here, $\mu^*$ is the Coulomb pseudopotential parameter, usually assigned a value of 0.13 [42]. Finally, the electron density in the Fermi energy level can also be estimated using the formula $N(E_F)=\frac{3}{\pi^2 k_B^2(1+\lambda_{ep})}\gamma$. The calculated $N(E_F)$ for the $Re_{0.3}Os_{0.3}Ta_{0.1}Nb_{0.1}W_{0.1}Zr_{0.1}$ is 0.89 states eV$^{-1}$ f.u.$^{-1}$. This almost agrees with the $N(E_F)$ values of other $\alpha$-Mn-type high-entropy superconducting alloys [33]. The angular independent gap function is expressed as $\Delta(T) = \alpha/\alpha_{BCS}\Delta_{BCS}(T)$ in the $\alpha$-model, where $\alpha_{BCS}$ = 1.76 denotes the weak-coupling gap ratio. The superconducting gap values can be obtained, with coupling strength $2\Delta_0/k_BT_c$ being 3.49 for $Re_{0.3}Os_{0.3}Ta_{0.1}Nb_{0.1}W_{0.1}Zr_{0.1}$. The relevant superconducting parameters of all samples are summarized in **Table 1**.

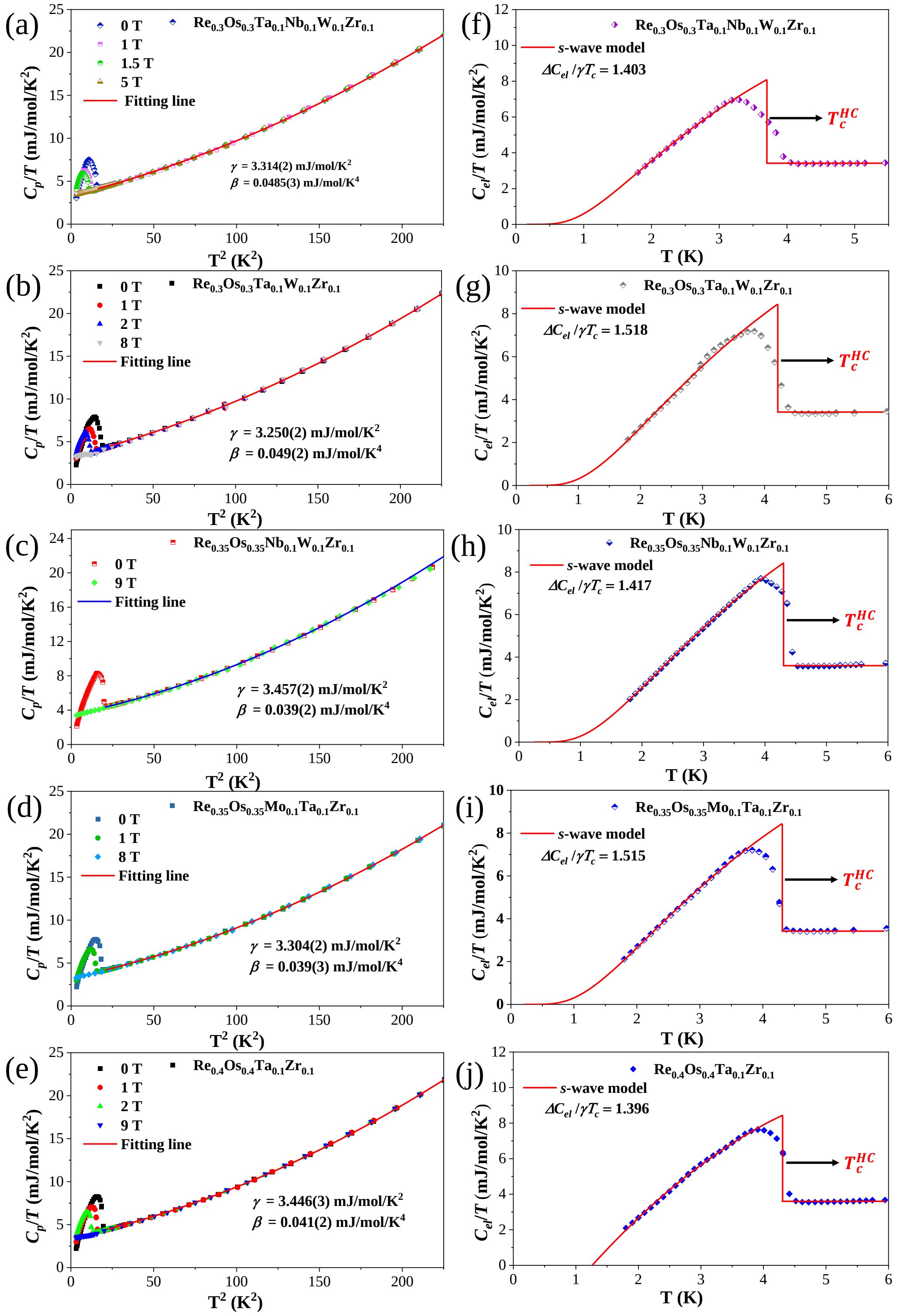

**Figure 5.** (a),(c),(e),(g), (i) The specific heat coefficient $C_p$/T as a function of $T^2$. (b), (d), (f), (h), (j) The electron contribution to specific heat as a function of $C_{el}$/T.

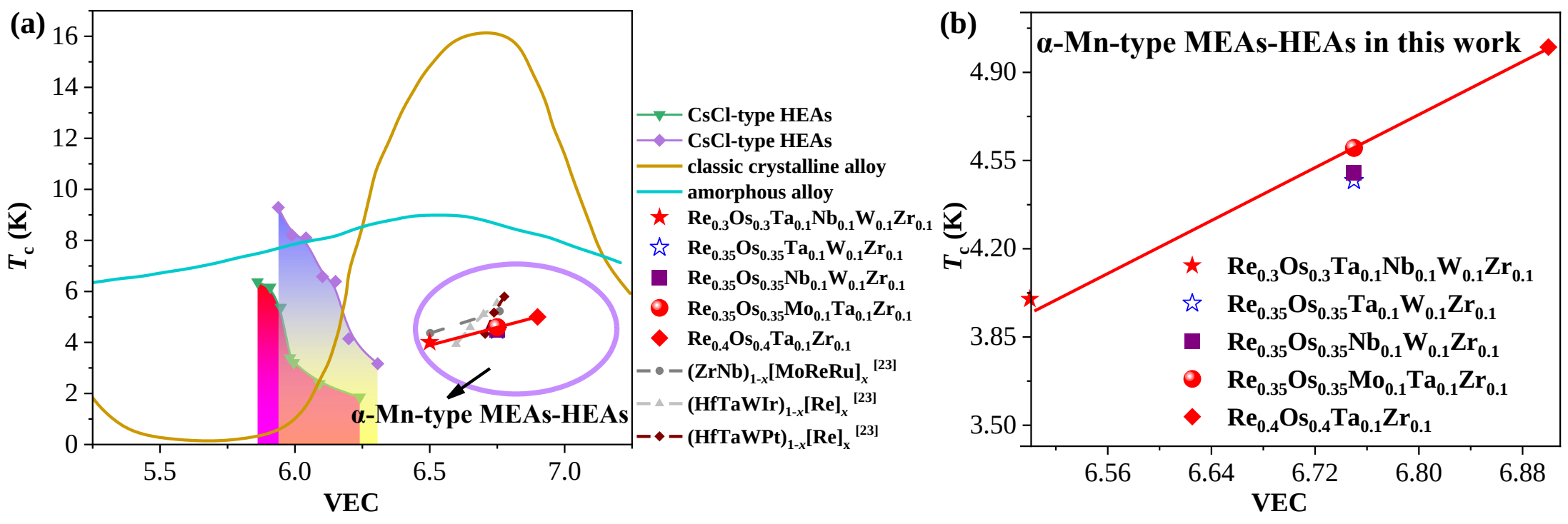

**Figure 6.** (a) Dependence of the VEC of superconductors on the $T_c$s. (b) The trend lines of the VEC dependence on the $T_c$s of $\alpha$-Mn-type samples in this paper.

The phase stability of crystalline solid solutions is significantly influenced by the parameter known as the valence electron count (VEC), particularly in alloys with comparable atomic size [13, 22]. A key step in alloy design involves calculating the VEC for each alloy, which contributes to the synthesis of superconducting materials. Our study summarizes the relationship between the $T_c$ and VEC for some selected $\alpha$-Mn-type, CsCl-type, and HCP-type compounds, as well as crystalline transition metals and amorphous vapor-deposited thin films. When comparing crystalline binary alloys and amorphous materials with the same electron count, we observe a positive effect of crystallinity on $T_c$. This behavior may stem from the ambiguity in electronic energy ($E$) and electronic wave vector ($k$), leading to an increase in disorder-induced electron state density [22]. Based on these findings, we speculate that improved crystallinity contributes to enhancing the $T_c$ of alloys. Furthermore, in CsCl-type superconductors, we have found a correlation between VEC and $T_c$, where the maximum $T_c$ reaches 9.3 K at a VEC value of around 5.9. However, the more complex $\alpha$-Mn-type MEA-HEA superconductors do not seem to confirm the overall trends observed in the other two systems. We compare the $T_c$ and VEC of crystalline transition metals and their alloys, amorphous vapor deposition films, CsCl-type HEAs and $\alpha$-Mn-type HEAs superconductors to summarize the relationship between them in **Figure 6a**. The trend lines of the VEC-dependent $T_c$ of $\alpha$-Mn-structure samples in this paper in **Figure 6b**. There is a certain regularity in the relationship between $T_c$ and VEC of

$Re_{0.3}Os_{0.3}Ta_{0.1}Nb_{0.1}W_{0.1}Zr_{0.1}$, $Re_{0.35}Os_{0.35}Ta_{0.1}W_{0.1}Zr_{0.1}$, $Re_{0.35}Os_{0.35}Nb_{0.1}W_{0.1}Zr_{0.1}$, $Re_{0.35}Os_{0.35}Mo_{0.1}Ta_{0.1}Zr_{0.1}$ and $Re_{0.4}Os_{0.4}Ta_{0.1}Zr_{0.1}$. Within a certain range, when the value of VEC increases, $T_c$ will also increase. Temperature variations induced by atomic substitution, especially through changes in VEC, significantly impact the performance of HEA superconductors. Therefore, in the exploration of novel superconducting materials through prediction, design, and synthesis, we believe that the relationship between VEC and $T_c$ obtained in this work could be the guide for synthesizing new superconductors.

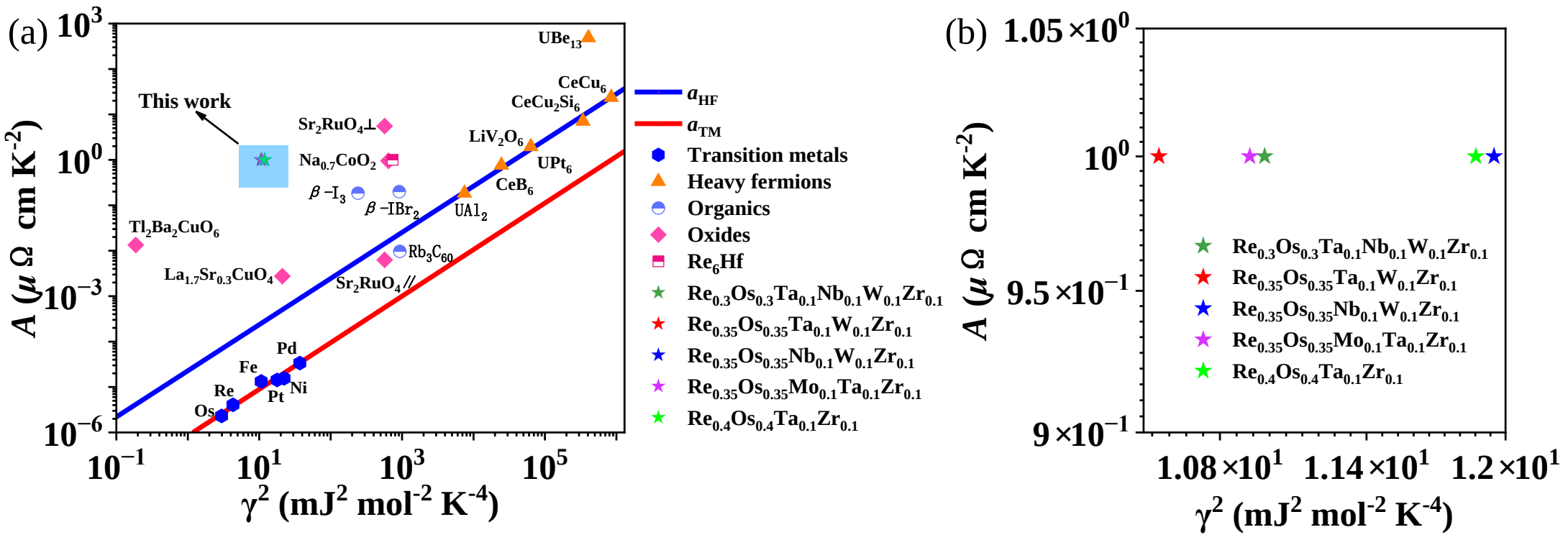

**Figure 7.** (a) The standard Kadowaki-Woods plot. The relationship between coefficient *A* and $\gamma^2$ for different compounds. (b) The relationship between coefficient *A* and $\gamma^2$ for $\alpha$-Mn structured superconductors in this paper.

In the field of condensed matter physics, based on the Fermi-liquid theory [43], the contribution of electrons to the heat capacity of a material follows a specific law with the expression $C_{el}(T) = \gamma T$, where $C_{el}(T)$ characterizes the heat capacity contributed by the electrons, $\gamma$ is the Sommerfeld coefficient, and $T$ represents the temperature. Meanwhile, the temperature-dependent resistivity of the material can be described as $\rho(T) = \rho_0 + AT^2$, where $\rho_0$ is the temperature-independent residual resistivity, and $A$ is the coefficient associated with electron-electron scattering. When electron-electron scattering, which dominates the quadratic term $AT^2$ in the resistivity equation, dominates the scattering process over electron-phonon scattering, the above temperature dependence of heat capacity and resistivity can be observed experimentally.

The Kadowaki-Woods ratio (KWR) is an important analytical tool to investigate the intrinsic relationship between the electron-electron scattering rate and the electron mass renormalization by comparing the temperature-dependent properties of the resistivity and heat capacity. The electron mass renormalization is a key indicator of the strength of the electron-electron correlation. The relationship $A/\gamma^2 \approx a_{TM} = 0.4$ μΩ cm mol$^2$ K$^2$ J$^{-2}$ = $0.04a_0$ is found for transition metals, and it is noteworthy that this relationship holds even when the value of $\gamma^2$ varies by up to an order of magnitude in the sample of transition metal materials studied. In contrast, $A/\gamma^2 \approx a_{HF} = 10$ μΩ cm mol$^2$ K$^2$ J$^{-2}$ = $a_0$ in numerous heavy fermionic material systems. In this research work, $Re_{0.35}Os_{0.35}Ta_{0.1}W_{0.1}Zr_{0.1}$ is selected as a typical representative material and its $\rho(T)$ data are fitted and analyzed in **Figure S3**. The coefficient $A = 6.39 \times 10^{-5}$ μΩ cm K$^{-2}$ as well as the residual resistivity $\rho_0 = 205$ μΩ cm were obtained. **Figure 7a** illustrates the relationship between the coefficient $A$ and $\gamma^2$ in different compounds of a variety of transition metals, heavy fermions, and other MEAs-HEAs [4, 12, 43-46]. Further, when $\gamma^2$ =10.56 mJ$^2$ mol$^{-2}$ K$^{-4}$ is used, the KWR value of $Re_{0.35}Os_{0.35}Ta_{0.1}W_{0.1}Zr_{0.1}$ HEA is calculated to be $6.05 \times 10^{-6}$ μΩ cm mol$^2$ K$^2$ mJ$^{-2}$. In addition, the relationship between the coefficient A and the Sommerfeld coefficient $\gamma^2$ for $\alpha$-Mn-structure superconductors, which are investigated in this paper, is presented in **Figure 7b**. Various mechanisms have been proposed to explain the variability of KWR values between different materials. It has been shown that the variation of KWR values depends on several factors such as material anisotropy, Fermi surface topology, carrier concentration, and magnetic correlation [43]. Compared with conventional transition metals, $\alpha$-Mn-structure MEAs-HEAs exhibit higher KWR values, a phenomenon suggesting that the alloys have richer dynamic properties. An in-depth analysis of the causes could be attributed to the disordered structure within the material or impurity-induced scattering, or it could be the result of electron-phonon interactions.

## Conclusion

In summary, we have synthesized previously unreported medium-/high-entropy alloys (MEAs-HEAs) $Re_{0.3}Os_{0.3}Ta_{0.1}Nb_{0.1}W_{0.1}Zr_{0.1}$, $Re_{0.35}Os_{0.35}Ta_{0.1}W_{0.1}Zr_{0.1}$, $Re_{0.35}Os_{0.35}Nb_{0.1}W_{0.1}Zr_{0.1}$, $Re_{0.35}Os_{0.35}Mo_{0.1}Ta_{0.1}Zr_{0.1}$, and $Re_{0.4}Os_{0.4}Ta_{0.1}Zr_{0.1}$ in the quaternary and quintuple system by arc melting of the 4*d* and 5*d* transition metals, with $T_{cs}$ of 4.0(3) K, 4.4(5) K, 4.5(2) K, 4.6(3) K, and 5.0(2) K, respectively. All these materials adopt a non-centrosymmetric (NC) $\alpha$-Mn-type structure. The presence of a large amount of Re seems to be important for the stability of the $\alpha$-Mn structure. We have achieved an increase in $T_c$ by increasing the VEC of NC $\alpha$-Mn structure HEA superconductors by varying the number of elements and changing the elemental species within a certain range. Specific-heat measurements show that all five alloys have an isotropic *s*-wave superconducting gap. The value of the superconducting gap $2\Delta_0/k_BT_c$ is about 3.5, and the value of the electron-phonon coupling strength $\lambda_{ep}$ is about 0.6, which might indicate the electron-phonon coupling mechanism of these superconductors. Furthermore, the $\alpha$-Mn-type MEAs-HEAs exhibit strong electron correlations comparable to those of typical heavy-fermion superconductors, a property revealed by the high value of KWR. These results render the $\alpha$-Mn-type MEAs-HEAs a new paradigm for studying the interaction between *s*-wave superconductivity and strong electron correlations. Therefore, this finding provides insight into the discovery of new $\alpha$-Mn-type MEA-HEA superconductors and provides new material platforms for investigating the effects of NC and high disorder on the superconducting mechanism.

**Notes**

The authors declare no competing financial interest

**Acknowledgment**

This work is funded by the Natural Science Foundation of China (No. 12274471, 12404165, 11922415), Guangzhou Science and Technology Programme (No. 2024A04J6415), the State Key Laboratory of Optoelectronic Materials and Technologies (Sun Yat-Sen University, No. OEMT-2024-ZRC-02), Key Laboratory of Magnetoelectric Physics and Devices of Guangdong Province (Grant No.

2022B1212010008), Research Center for Magnetoelectric Physics of Guangdong Province (2024B0303390001). Lingyong Zeng was thankful for the Postdoctoral Fellowship Program of CPSF (GZC20233299), and Fundamental Research Funds for the Central Universities, Sun Yat-sen University (29000-31610058).

# Supporting Information

## Superconductivity in the Medium-Entropy/High-Entropy Re-based Alloys with a Non-Centrosymmetric $\alpha$-Mn Lattice

*Kuan Li*[a], *Longfu Li*[a], *Lingyong Zeng*[a,b], *Yucheng Li*[a], *Rui Chen*[a], *Peifeng Yu*[a], *Kangwang Wang*[a], *Zaichen Xiang*[a], *Tian Shang*[c], *Huixia Luo*[adef*]

[a]School of Materials Science and Engineering, Sun Yat-sen University, No. 135, Xingang Xi Road, Guangzhou, 510275, P. R. China

[b]Device Physics of Complex Materials, Zernike Institute for Advanced Materials, University of Groningen, Nijenborgh 4, 9747 AG Groningen, The Netherlands

[c]Key Laboratory of Polar Materials and Devices (MOE), School of Physics and Electronic Science, East China Normal University, Shanghai 200241, China

[d]State Key Laboratory of Optoelectronic Materials and Technologies, Sun Yat-sen University, No. 135, Xingang Xi Road, Guangzhou, 510275, P. R. China

[e]Key Lab of Polymer Composite & Functional Materials, Sun Yat-sen University, No. 135, Xingang Xi Road, Guangzhou, 510275, P. R. China

[f]Guangdong Provincial Key Laboratory of Magnetoelectric Physics and Devices, Sun Yat-sen University, No. 135, Xingang Xi Road, Guangzhou, 510275, P. R. China

*Corresponding author/authors complete details (Telephone; E-mail:) (+86)-2039386124; E-mail address: luohx7@mail.sysu.edu.cn;

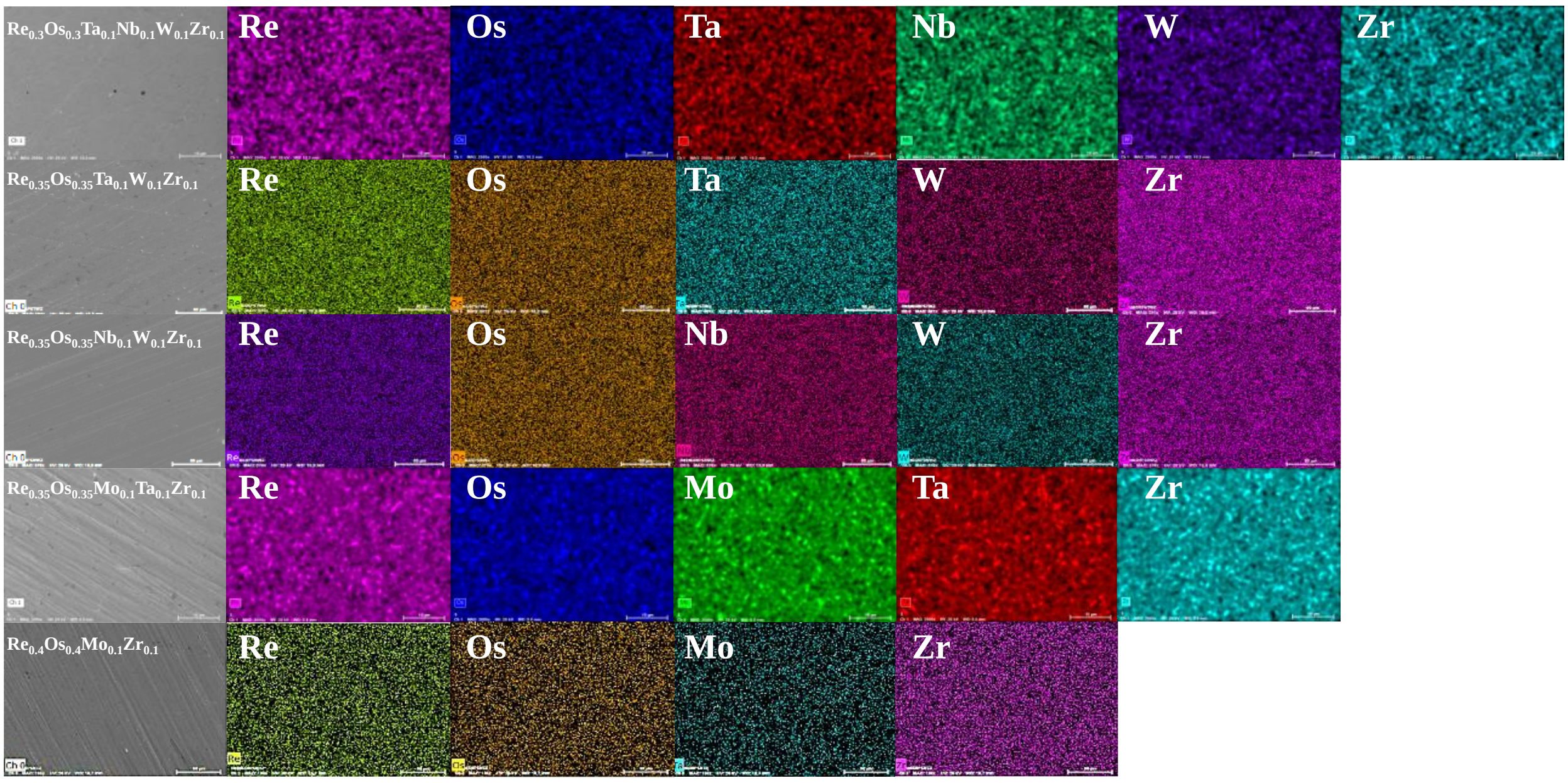

**Figure S1.** SEM images and EDX elemental mappings of $Re_{0.3}Os_{0.3}Ta_{0.1}Nb_{0.1}W_{0.1}Zr_{0.1}$, $Re_{0.35}Os_{0.35}Ta_{0.1}W_{0.1}Zr_{0.1}$, $Re_{0.35}Os_{0.35}Nb_{0.1}W_{0.1}Zr_{0.1}$, $Re_{0.35}Os_{0.35}Mo_{0.1}Ta_{0.1}Zr_{0.1}$, and $Re_{0.4}Os_{0.4}Ta_{0.1}Zr_{0.1}$ superconductors.

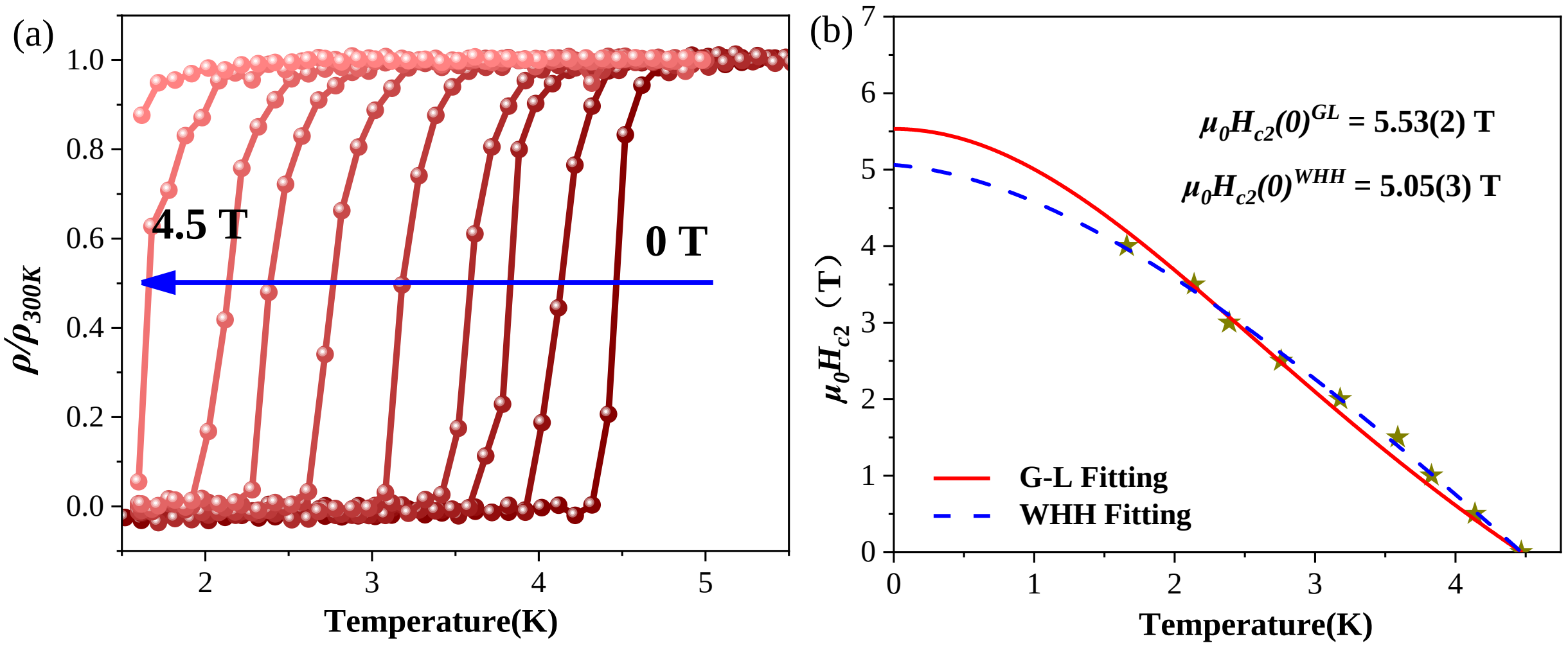

**Figure S2.** (a) The resistive transition under different magnetic fields. (b) The $H_{c2}$-T phase diagram of $Re_{0.35}Os_{0.35}Ta_{0.1}W_{0.1}Zr_{0.1}$ superconductor.

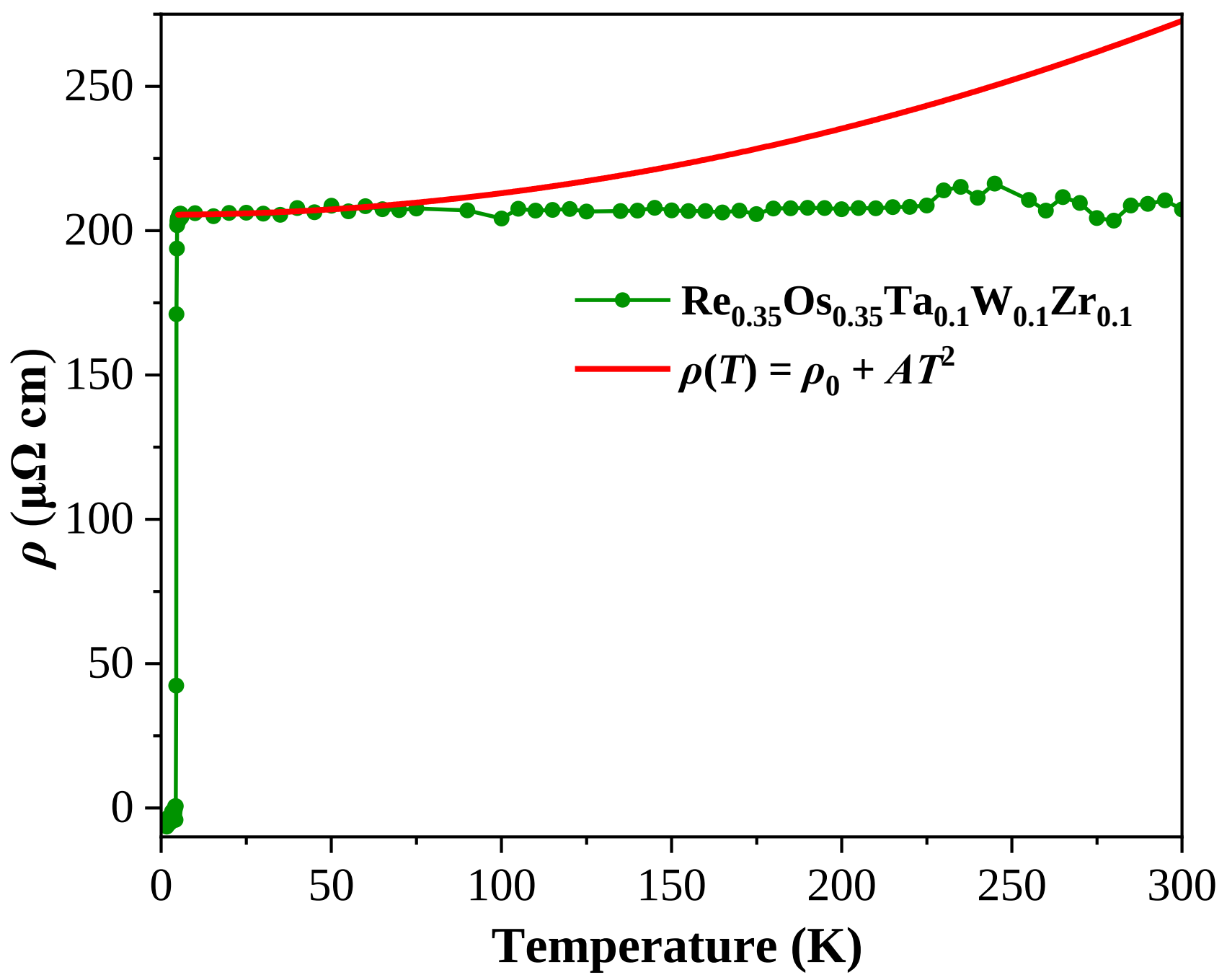

**Figure S3.** (a) . Temperature dependence of resistivity at zero field, where the data above $T_c$ up to 75 K is well fitted by power law: $\rho(T) = \rho_0 + AT^2$, shown by red solid line.

**Table S1.** The XRD fitting parameters and unit cell parameters of $Re_{0.3}Os_{0.3}Ta_{0.1}Nb_{0.1}W_{0.1}Zr_{0.1}$, $Re_{0.35}Os_{0.35}Ta_{0.1}W_{0.1}Zr_{0.1}$, $Re_{0.35}Os_{0.35}Nb_{0.1}W_{0.1}Zr_{0.1}$, $Re_{0.35}Os_{0.35}Mo_{0.1}Ta_{0.1}Zr_{0.1}$, and $Re_{0.4}Os_{0.4}Ta_{0.1}Zr_{0.1}$ superconductors.

| Samples | lattice parameter (Å) | $R_{wp}$ | $R_p$ | $R_e$ | $\chi^2$ | $\Delta S_{mix}$ |
|---|---|---|---|---|---|---|
| $Re_{0.3}Os_{0.3}Ta_{0.1}Nb_{0.1}W_{0.1}Zr_{0.1}$ | $a = b = c$ = 9.589(2) | 7.42 | 5.22 | 2.84 | 6.8033 | 1.64R |
| $Re_{0.35}Os_{0.35}Ta_{0.1}W_{0.1}Zr_{0.1}$ | $a = b = c$ = 9.618(6) | 5.53 | 3.87 | 2.52 | 4.8081 | 1.43R |
| $Re_{0.35}Os_{0.35}Nb_{0.1}W_{0.1}Zr_{0.1}$ | $a = b = c$ = 9.661(13) | 5.97 | 4.45 | 2.57 | 5.3917 | 1.43R |
| $Re_{0.35}Os_{0.35}Mo_{0.1}Ta_{0.1}Zr_{0.1}$ | $a = b = c$ = 9.551(13) | 5.43 | 4.07 | 2.80 | 3.7604 | 1.43R |
| $Re_{0.4}Os_{0.4}Ta_{0.1}Zr_{0.1}$ | $a = b = c$ = 9.606(3) | 4.77 | 3.69 | 2.48 | 3.6892 | 1.19R |

**Table S2.** The atomic parameters of a representative sample of $Re_{0.4}Os_{0.4}Ta_{0.1}Zr_{0.1}$ superconductor.

| | | Structure | | Cubic |
|---|---|---|---|---|
| | | Space group | | $I\bar{4}3m$ |
| | | Lattice parameters $a$ (Å) | | 9.6408 |
| Atom | *x* | *y* | *z* | Occupancy |
| Os1 | 0.000 | 0.000 | 0.000 | 0.4 |
| Os2 | 0.316 | 0.316 | 0.316 | 0.4 |
| Os3 | 0.091 | 0.091 | 0.091 | 0.4 |
| Os4 | 0.359 | 0.359 | 0.359 | 0.4 |
| Ta1 | 0.000 | 0.000 | 0.000 | 0.1 |
| Ta2 | 0.359 | 0.359 | 0.359 | 0.1 |
| Ta3 | 0.316 | 0.316 | 0.316 | 0.1 |
| Ta4 | 0.091 | 0.091 | 0.091 | 0.1 |
| Re1 | 0.000 | 0.000 | 0.000 | 0.4 |
| Re2 | 0.316 | 0.316 | 0.316 | 0.4 |
| Re3 | 0.091 | 0.091 | 0.091 | 0.4 |
| Re4 | 0.359 | 0.359 | 0.359 | 0.4 |
| Zr1 | 0.000 | 0.000 | 0.000 | 0.1 |
| Zr2 | 0.316 | 0.316 | 0.316 | 0.1 |
| Zr3 | 0.091 | 0.091 | 0.091 | 0.1 |
| Zr4 | 0.359 | 0.359 | 0.359 | 0.1 |

**Table S3.** The element ratios of Samples based on multiple EDS measurements.

| Sample \ Element ratio | Re | Os | Ta | W | Zr | Nb | Mo |
|---|---|---|---|---|---|---|---|
| $Re_{0.3}Os_{0.3}Ta_{0.1}Nb_{0.1}W_{0.1}Zr_{0.1}$ | 0.32 | 0.31 | 0.12 | 0.09 | 0.08 | 0.08 | / |
| $Re_{0.35}Os_{0.35}Ta_{0.1}W_{0.1}Zr_{0.1}$ | 0.36 | 0.34 | 0.09 | 0.11 | 0.10 | / | / |
| $Re_{0.35}Os_{0.35}Nb_{0.1}W_{0.1}Zr_{0.1}$ | 0.35 | 0.34 | / | 0.12 | 0.11 | 0.08 | |
| $Re_{0.35}Os_{0.35}Mo_{0.1}Ta_{0.1}Zr_{0.1}$ | 0.34 | 0.36 | 0.10 | / | 0.09 | / | 0.11 |
| $Re_{0.4}Os_{0.4}Ta_{0.1}Zr_{0.1}$ | 0.42 | 0.38 | 0.09 | | 0.11 | | |